\documentclass[superscriptaddress,twocolumn]{revtex4-2}

\usepackage[english]{babel}
\usepackage[utf8]{inputenc}
\usepackage{amsmath}
\usepackage{amsfonts}
\usepackage{amssymb}
\usepackage{physics}
\usepackage{bbold}
\usepackage{bbm} % bold math
\usepackage{graphicx}
\usepackage[normalem]{ulem}
\usepackage{color}
\usepackage{hyperref}
\usepackage[nameinlink]{cleveref}
\usepackage{mathtools}
\usepackage{changepage}
\usepackage{setspace}
\usepackage{verbatim}
\usepackage{tcolorbox}
\usepackage{float}

\hypersetup{
	colorlinks,
	linkcolor={red!40!black},
	citecolor={blue!60!black},
	urlcolor={blue!50!black}
	}

\begin{document}

\title{Dressing the Lorentz atom}

\author{Stephen M. Barnett}
\affiliation{School of Physics and Astronomy, University of Glasgow, Glasgow G12 8QQ, UK}

\author{James D. Cresser}
\affiliation{Department of Physics and Astronomy, University of Exeter, Stocker Road, Exeter EX4 4QL, UK}
\affiliation{School of Physics and Astronomy, University of Glasgow, Glasgow, G12 8QQ, UK}
\affiliation{Department of Physics and Astronomy, Macquarie University, 2109 NSW, Australia}

\author{Sarah Croke}
\affiliation{School of Physics and Astronomy, University of Glasgow, Glasgow G12 8QQ, UK}
% \today

%%%%%%%%%%%%%%%%%%%%%%%%%%%%%
%%%%%%%%%%%%%%%%%%%%%%%%%%%%%
%%%%%%%%%%%%%%%%%%%%%%%%%%%%%

\begin{abstract}
We investigate the effects of the electromagnetic vacuum field on a harmonically bound electron.  We show that in the electric-dipole approximation the model atom couples only to an effective one-dimensional electric field. In a simplified form, in which the problem is reduced to a single spatial dimension, we determine, analytically, the form of the ground state and discuss the significance of this.
\end{abstract}
\maketitle

\section{Introduction}
\label{Sect1}

In quantum electrodynamics (QED) a charged particle, such as an electron, interacts with its surrounding electromagnetic field.  This is true even for an electron in the vacuum \cite{Schwinger,Feynman}.  Some important consequences of this interaction, such as the Lamb shift and the anomalous magnetic moment of the electron, have been investigated in great detail \cite{BetheSalpeter}.  At low energies, we can apply non-relativistic methods to address these problems and this approach, pioneered and championed by Power and Thirunamachandran in London and by Persico, Compagno and Passante in Palermo, has revealed a wealth of quantum effects, notably in the interactions between diverse molecules \cite{EdwinBook,ThiruBook,PersicoBook,Salam}.  The simplest example of this is the Casimir-Polder interaction in which the familiar van der Waals force of attraction between molecules at short distances \cite{vdW} is replaced by a weaker, and intrinsically quantum, force at larger distances \cite{EdwinBook,PersicoBook,CasimirP,MilonniBook}. It is often the case, particularly in quantum optics, that such effects can be ignored in favour of much stronger resonant effects involving the emission and absorption of photons in near-resonant transitions.  Even in such circumstances, however, non-resonant virtual transitions can have a role as, for example, in the treatment of linear polarisabilities
\cite{LoudonJPB,BermanPRA,MilonniPRA08}.

The expression `dressed state' has been used in two distinct, but related, ways and we should be clear as to its usage here.  To make the point, we can do no better than to quote directly from the opening of a paper by Compagno, Passante and Persico \cite{ContempPhys} (with some updated references):

{\em The expression `dressed atom' in non-relativistic QED is used with reference to diverse physical situations which, broadly speaking, differ with respect to the properties of the electromagnetic field with which the atom is assumed to interact.  Historically, the earliest use of this expression was in connection with an atom in the presence of real photons, such as those produced by an external source of electromagnetic radiation \cite{CohenTKastler} and the corresponding notion has found wide and well-known applications mainly (but not only) in quantum optics \cite{CohenTBook,PaulR,Ficek,Agarwal}.  In contrast here, focussing on the situation where one has a ground-state bare atom interacting with the vacuum electromagnetic field, we take the total atom-field system to be in its lowest possible energy state.  Thus the zero-point quantum fluctuations of the field can only induce virtual absorption and re-emission processes by the atom.  Since these processes take place continuously, they create a cloud of virtual photons around the bare atom.  The complex object (atom + cloud of virtual photons) is what we mean by `dressed atom' in this paper
\cite{PersicoBook}.}

In treating interactions between molecules or atoms it suffices, often, to employ perturbation theory \cite{EdwinBook,ThiruBook,PersicoBook,Salam}.  This results in energy-level shifts and hence forces that depend only on the distance between the molecules and a few simple parameters, such as ground-state polarizabilities.  The level shifts can be pictured as arising from the exchange of virtual photons arising from off-resonance transitions in which an electron can make a transition to a higher energy level while simultaneously emitting a photon.  This photon can then be absorbed, albeit briefly, by a second atom or molecule. The term virtual is, at least potentially, misleading in that the virtual photon is a regular excitation of the quantized electromagnetic field but the energy non-conserving interaction that created it means that the photon can only be short lived.  There is, nevertheless, a cloud of such virtual photons around all charged particles and, of particular interest to non-relativistic QED, around bound systems of such charges, including atoms and molecules.  This cloud, moreover, has features of its atomic or molecular source imprinted upon it and it is this dressing of the atom or molecule that controls the interactions between them \cite{PersicoBook,Passante85,CompagnoPS3,CompagnoAPP,CompagnoRev}.  Furthermore, abrupt changes in the environment surrounding the source can have a correspondingly dramatic effect on the cloud, including changes between dressing and half-dressing
\cite{PersicoBook,CompagnoRev,Persico87,CompagnoPRA88,CompagnoPS1,CompagnoPS2}.
A rather direct example of this occurs for an atom in an excited state placed in a cavity that is too small to allow spontaneous emission to take place, but then has one of the mirrors suddenly removed \cite{Fearn}.

The complicated nature of atoms, and perhaps especially of molecules, often enforces the use of perturbation theory in the description of the virtual photons dressing the ground state.  For this reason it may be helpful, even if only as an aid to physical reasoning, to treat exactly a model system.  To this end we consider dressing the Lorentz model of an atom in which a single electron is considered to be harmonically bound about the origin.  We find that, at least in a simplified one-dimensional form, it is possible to determine exactly the form of the ground state of the bound electron, dressed by the surrounding vacuum field.

\section{The Lorentz atom}
\label{Sect2}

The Lorentz model treats an electron as a classical particle harmonically bound to a fixed point in space, which we can conveniently take as the origin of coordinates.  This electron will respond to an externally applied electric field and hence undergo forced, damped simple harmonic motion.  This motion is determined by the simple equation
\begin{equation}
\label{Eq2.1}
\frac{d^2}{dt^2}{\bf x} + \gamma\frac{d}{dt}{\bf x} + \Omega_0^2{\bf x} = -\frac{e}{m}{\bf E} \, ,
\end{equation}
where $-e$ and $m$ are, respectively, the charge and mass of the electron. Lorentz used this model to treat, successfully, the Zeeman effect \cite{Lorentz}, and it is widely employed in derivations of the dielectric constant for material media \cite{Jackson,Bleaney,Dressel}.  If the natural frequency, $\Omega_0$, is set to zero then we recover the Drude model \cite{Drude}, which is regularly employed in the description of metals and other conductors \cite{Ashcroft,Wooten}. We note that something like a harmonically trapped electron has been achieved in a Penning trap and used to make some of the most accurate measurements of fundamental properties of the electron \cite{Dehmelt,Ghosh}.

To investigate the effects of the vacuum field on a Lorentz-model atom, we require the quantum mechanical form of the evolution equation and, moreover, the Hamiltonian that underlies this.  The quantum form of the evolution equation is \cite{MilonniBook}
\begin{equation}
\label{Eq2.2}
\frac{d^2}{dt^2}\hat{\bf x} + \Omega_0^2\hat{{\bf x}} = -\frac{e}{m}\hat{\bf E} \, .
\end{equation}
There are important differences of form between this operator equation and the classical expression given above.  The most obvious of these is the omission of a damping term for the oscillator.  This does exist, but is hidden within the driving term proportional to the electric field operator.  This is the full field, including the radiation reaction term that, in conjunction with the vacuum fluctuations, gives rise to the damping \cite{MilonniBook}.  A second, more subtle feature, is that we take the form of the electric field operator at the origin rather than at the position of the electron.  This is the familiar electric dipole approximation \cite{QTL}, which can be justified as follows.  A typical length scale for the Lorentz atom is given by the width of the ground state of the oscillator, which is $\Delta x = (\hbar/2m\Omega_0)^{1/2}$, and the wavelength of resonant radiation at the same frequency is $\lambda_0 = 2\pi c/\Omega_0$.  For the dipole approximation to hold we require $\lambda_0 \gg \Delta x$.  This provides a restriction on the angular frequency in the form
\begin{equation}
\label{Eq2.3} 
\hbar\Omega_0 \ll 8\pi^2mc^2  ,
\end{equation}
the upper limit of which lies in the ultra-relativistic regime.  It follows that the dipole approximation is an excellent one for the non-relativistic domain in which we are working.

Our aim is to determine the form of the ground state of the Lorentz atom and to determine this we require the form of the Hamiltonian that gives rise to the evolution equation (\ref{Eq2.2}).  It is straightforward to show that the required form is
\begin{equation}
\label{Eq2.4}
\hat{H} = \frac{\hat{p}^2}{2m} + \frac{1}{2}m\Omega_0^2\hat{r}^2 + e\hat{\bf r}\cdot\hat{\bf E}(0) + \hat{H}_{\rm rad} \, ,
\end{equation}
where $\hat{H}_{\rm rad}$ is the Hamiltonian for the free electromagnetic field.  We note that if we replace the potential term, $\frac{1}{2}m\Omega_0^2\hat{r}^2$, by the Coulomb potential of a proton, we recover the lowest order form of the multipolar Hamiltonian for the interaction between an atom and the electromagnetic field.

\section{Dressing the Lorentz atom: what to expect}
\label{Sect3}

The Lorentz-atom Hamiltonian has the form of three harmonic oscillators, one for each of the three cartesian directions, coupled to the electric field.  The latter also has the form of a number of harmonic oscillators, one for each mode of the field.  It follows that the equations of motion for the annihilation and creation operators for each of the electron motion and field oscillators are linear and so, in principle, the associated coupling matrix can be diagonalised.  In practice, however, this is difficult to achieve analytically and in closed form, but a simplified version of the problem is tractable.  Before presenting this, it is instructive to treat a yet simpler problem of just two coupled mechanical oscillators
with a Hamiltonian in the form 
\begin{equation}
\label{Eq3.1}
\hat{H} = \frac{1}{2m}\left(\hat{p}_1^2 + \hat{p}_2^2\right) 
+ \frac{1}{2}m\Omega_0^2\left(\hat{x}_1^2 + \hat{x}_2^2 - 2g\hat{x}_1\hat{x}_2\right)  ,
\end{equation}
where the parameter $g$ determines the strength of the coupling between the two oscillators; for the two oscillators to execute coupled simple harmonic motion we require $|g|<1$.

It is straightforward to diagonalise our Hamiltonian (\ref{Eq3.1}) by introducing the normal-mode operators
\begin{align}
\label{Eq3.2}
\hat{x}_\pm = \frac{1}{\sqrt{2}}\left(\hat{x}_1 \pm \hat{x}_2\right)  \nonumber \\
\hat{p}_\pm = \frac{1}{\sqrt{2}}\left(\hat{p}_1 \pm \hat{p}_2\right) ,
\end{align}
so that our Hamiltonian becomes
\begin{equation}
\label{Eq3.3}
\hat{H} = \frac{\hat{p}_+^2}{2m} + \frac{1}{2}m\Omega_0^2(1-g)\hat{x}_+^2
+ \frac{\hat{p}_-^2}{2m} + \frac{1}{2}m\Omega_0^2(1+g)\hat{x}_-^2 \, .
\end{equation}
It then follows that the ground state is simply the product of the uncoupled ground states for the two normal modes:
\begin{equation}
\label{Eq3.4}
\hat{H}|0\rangle_+|0\rangle_- = \frac{1}{2}\hbar\Omega_0\left(\sqrt{1-g} + \sqrt{1+g}\right)
|0\rangle_+|0\rangle_- \, .
\end{equation}
This eigenenergy is clearly less than the ground-state energy, $\hbar\Omega_0$, for the two uncoupled oscillators.  The mean energy for oscillator 1 alone, as determined from the uncoupled Hamiltonian, is greater than that for the uncoupled oscillator:
\begin{widetext}
	\begin{equation}
\label{Eq3.5}
{}_+\langle 0|{}_-\langle 0|\frac{\hat{p}_1^2}{2m} + \frac{1}{2}m\Omega_0^2\hat{x}_1^2|0\rangle_+|0\rangle_- 
=
\frac{1}{8}\hbar\Omega_0\left(\sqrt{1-g} + \sqrt{1+g} +\frac{1}{\sqrt{1-g}} + \frac{1}{\sqrt{1+g}}\right) >\frac{1}{2}\hbar\Omega_0 \, .
\end{equation}
\end{widetext}
Finally, we note that the ground state of the coupled oscillators is an entangled state of oscillators 1 and 2.  The easiest way to demonstrate this is to note the correlations that exist between the positions and momenta of the oscillators:
\begin{align}
\label{Eq3.6}
{}_+\langle 0|{}_-\langle 0|\hat{x}_1\hat{x}_2|0\rangle_+|0\rangle_- &= 
\frac{\hbar}{4m\Omega_0}\left(\frac{1}{\sqrt{1-g}} - \frac{1}{\sqrt{1+g}}\right)  \nonumber \\
{}_+\langle 0|{}_-\langle 0|\hat{p}_1\hat{p}_2|0\rangle_+|0\rangle_- &=
\frac{\hbar m\Omega_0}{4}\left(\sqrt{1-g} - \sqrt{1+g}\right) .
\end{align}
These mean that the positions for the two oscillators are correlated but that the two momenta are anti-correlated. The cross-correlations between the position of one oscillator and the momentum of the other are zero:
\begin{align}
\label{Eq3.7}
{}_+\langle 0|{}_-\langle 0|\hat{x}_1\hat{p}_2|0\rangle_+|0\rangle_- 
&= {}_+\langle 0|{}_-\langle 0|\hat{p}_1\hat{x}_2|0\rangle_+|0\rangle_- \notag\\
&= 0.
\end{align}

The features that we can expect to find for the dressed Lorentz atom are, therefore, an entangled ground state with correlations between the electron motion and the surrounding virtual photons and a decrease in the ground state energy combined with
an increased energy associated with the electron motion alone. 

\section{Dressing the one-dimensional Lorentz atom}
\label{Sect4}

We seek an expression for the dressed ground state of the Lorentz atom and to use this to better understand the properties of dressed atoms.  Realising this in three dimensions is mathematically challenging and so, very much in the interests of simplicity, we reduce the problem of diagonalising the Lorentz-atom Hamiltonian to a single spatial dimension. There is, also, a physical case for pursuing this as an electron in a Penning trap may experience an axial trapping frequency that is very different from those associated with the other two spatial dimensions and so can be treated separately \cite{Ghosh}.  We note, also, that there exists a substantial body of work on the one-dimensional hydrogen atom \cite{LoudonAmJ,LoudonProc} but the one-dimensional Lorentz atom is significantly easier to treat.
\begin{widetext}
For our one-dimensional model, the Lorentz-atom Hamiltonian (\ref{Eq2.4}) becomes
\begin{equation}
\label{Eq4.1}
\hat{H} = \frac{\hat{p}_z^2}{2m} + \frac{1}{2}m\Omega_0^2\hat{z}^2 + e\hat{z}\hat{E}_z(0) + \hat{H}_{\rm rad} \, .
\end{equation}
It is straightforward to show that the $z$-component of the electric field operator can be expressed in the form of a frequency integral over suitably weighted continuum annihilation and creation operators:

	\begin{equation}
\label{Eq4.2}
\hat{H} = \frac{\hat{p}_z^2}{2m} + \frac{1}{2}m\Omega_0^2\hat{z}^2 + 
e\hat{z}\left(\frac{\hbar}{6\pi^2\varepsilon_0c^3}\right)^{1/2}\int_0^\infty \omega^{3/2}  \left(\hat{b}(\omega) + \hat{b}^\dagger(\omega)\right)d\omega
+ \int_0^\infty \hbar\omega\hat{b}^\dag(\omega)\hat{b}(\omega)d\omega\, ,
\end{equation}
% \end{widetext}
where the annihilation and creation operators satisfy the familiar bosonic commutation relation 
\begin{equation}
\label{Eq4.3}
\left[\hat{b}(\omega),\hat{b}^\dagger(\omega')\right] = \delta(\omega-\omega') .
\end{equation}
It is convenient to write the Hamiltonian in terms of the familiar annihilation and creation operators for the atomic oscillator:
% \begin{widetext}
	\begin{equation}
\label{Eq4.4}
\hat{H} = \hbar\Omega_0\hat{a}^\dagger\hat{a} + 
\left(\hat{a} + \hat{a}^\dagger\right)
\left(\frac{\hbar^2e^2}{12\pi^2\varepsilon_0c^3m\Omega_0}\right)^{1/2}\int_0^\infty \omega^{3/2}  \left(\hat{b}(\omega) 
+ \hat{b}^\dagger(\omega)\right)d\omega
+ \int_0^\infty \hbar\omega\hat{b}^\dag(\omega)\hat{b}(\omega)d\omega\, ,
\end{equation}
% \end{widetext}
where we have removed the unimportant ground-state energy of the undressed atom.  We note that in this form the electron appears coupled to a single field mode (albeit a continuum one) for each frequency rather than the more usual field depending on the full wavevector ${\bf k}$.  That this is possible is a consequence of the electric dipole approximation.  Details of this derivation are given in Appendix \ref{AppA}.

It is convenient to write this Hamiltonian in the more general form
% \begin{widetext}
	\begin{equation}
\label{Eq4.5}
\hat{H} = \hbar\Omega_0\hat{a}^\dagger\hat{a} + 
\int_0^\infty \hbar\omega\hat{b}^\dag(\omega)\hat{b}(\omega)d\omega 
+ \int_0^\infty \frac{\hbar}{2}V(\omega) \left(\hat{a} + \hat{a}^\dagger\right) \left(\hat{b}(\omega) 
+ \hat{b}^\dagger(\omega)\right)d\omega \, ,
\end{equation}
\end{widetext}
where $V(\omega)$ is yet to be specified.  The reason for this is that there is a positivity constraint that limits those forms of the coupling for which diagonalization is possible \cite{HOpaper}.  Specifically, we require that
\begin{equation}
\label{Eq4.6}
\int_0^\infty d\omega\frac{V^2(\omega)}{\omega} < \Omega_0 \, .
\end{equation}
As it stands, for $V(\omega)$ in eq. (\ref{Eq4.4}), the integral in this inequality diverges and so, in order to proceed, we introduce a cutoff by writing
\begin{equation}
\label{Eq4.7}
V^2(\omega) = \omega^3e^{-\omega/\omega_c}
\left(\frac{e^2}{3\pi^2\varepsilon_0c^3m\Omega_0}\right) \, .
\end{equation}
This restriction on the coupling strength is analogous to that on the value of the coupling constant $g$ for two coupled oscillators, as treated in the preceding section.  It is physically reasonable because we are performing a non-relativistic calculation and working within the electric dipole approximation, and both of these features will fail at sufficiently large frequencies.  Evaluating the integral leads to the bound
\begin{equation}
\label{Eq4.8}
\omega_c < \left(\frac{3\pi^2\varepsilon_0c^3m}{2e^2}\Omega_0^2\right)^{1/3}
\approx \Omega_0^{2/3}\times 6 \times 10^7 s^{-1/3} ,
\end{equation}
with the (angular) frequencies expressed in inverse seconds.  It is reasonable to require, also, that the cutoff frequency is significantly greater than $\Omega_0$. These conditions are readily satisfied for systems of physical interest.  We find, however, that the qualitative properties of the dressed atom are largely independent of the form of the coupling to the field, as long as the condition (\ref{Eq4.7}) is satisfied.

Let us return to the problem of diagonalizing the Hamiltonian (\ref{Eq4.5}).  By diagonalization we mean finding a complete set of continuum annihilation operators $\hat{B}(\omega)$ with the property that
\begin{equation}
\label{Eq4.9}
\left[\hat{B}(\omega),\hat{H}\right] = \omega\hat{B}(\omega)
\end{equation}
for the continuum of frequencies greater than $0$.  The operators will then be complete if we can construct the operators $\hat{a}$, $\hat{a}^\dagger$, $\hat{b}(\omega')$ and $\hat{b}^\dagger(\omega')$ as linear combinations of $\hat{B}(\omega)$ and $\hat{B}^\dagger(\omega)$.  Determining the forms of $\hat{B}(\omega)$ and $\hat{B}^\dagger(\omega)$ is a subtle process and we refer the reader to the earlier literature for the details \cite{HOpaper,Bruno}, but give a brief discussion of this in Appendix \ref{AppB}.  If we write $\hat{B}(\omega)$ in the form of a superposition,
\begin{align}
\label{Eq4.10}
\hat{B}(\omega) &= \alpha(\omega)\hat{a} + \beta(\omega)\hat{a}^\dagger \notag\\
&\quad
+\int_0^\infty d\omega'[\gamma(\omega,\omega')\hat{b}(\omega') + \delta(\omega,\omega')\hat{b}^\dagger(\omega')]
\end{align}
then imposing the conditions (\ref{Eq4.9}) together with the commutator
$[\hat{B}(\omega),\hat{B}^\dagger(\omega')] = \delta(\omega-\omega')$ leads to the (exact) expressions 
\begin{align}
\label{Eq4.11}
\alpha(\omega) &=  \frac{\omega + \Omega_0}{\Omega_0V(\omega)}\left(\frac{1}{Y(\omega) - i\pi}\right) \nonumber \\
\beta(\omega) &=  \frac{\omega - \Omega_0}{\omega + \Omega_0}\alpha(\omega) \nonumber \\
\gamma(\omega,\omega') &=  \left(\frac{\mathbbmss{P}}{\omega-\omega'} + Y(\omega)\delta(\omega-\omega')\right)
V(\omega')\frac{\Omega_0}{\omega+\Omega_0}\alpha(\omega) \nonumber \\
\delta(\omega,\omega') &=  \left(\frac{1}{\omega+\omega'}\right)V(\omega')\frac{\Omega_0}{\omega+\Omega_0}\alpha(\omega)  ,
\end{align}
where $\mathbbmss{P}$ denotes a principal part and $Y(\omega)$ is the real function
\begin{widetext}
	\begin{equation}
\label{Eq4.12}
Y(\omega) = \frac{1}{V^2(\omega)}\left[\frac{2(\omega^2-\Omega_0^2)}{\Omega_0} -
\int_0^\infty d\omega' \left(\frac{\mathbbmss{P}}{\omega - \omega'} - \frac{1}{\omega + \omega'}\right)V^2(\omega')\right] .
\end{equation}
\end{widetext}
The true ground state of our one-dimensional Lorentz atom, which we denote by the ket $|0\rangle$, is the zero eigenvalue right eigenstate of all our dressed annihilation operators:
\begin{equation}
\label{Eq4.13}
\hat{B}(\omega)|0\rangle = 0 .
\end{equation}
It is clear, directly from the construction of the $\hat{B}(\omega)$, that this pure state is an entangled state of the bare atom and the surrounding electromagnetic field.  We turn next to examine the properties of this ground state.

\section{Properties of the dressed ground state}
\label{Sect5}
We start by emphasizing the point that the diagonalization of the Hamiltonian is exact and so the state $|0\rangle$ is the true ground state.  In principle, all of the properties of this state, including the excitations of the (bare) oscillator and the statistics of the surrounding (virtual) photon cloud can be determined.  Evaluating the necessary integrals, however, is both a subtle and a technically demanding task and so, in this necessarily brief paper, we concentrate on a qualitative presentation of some of the key properties.

The first thing to note is that the properties of our dressed Lorentz atom can be expressed, simply, in terms of a single function
\begin{equation}
\label{Eq5.1}
\pi(\omega) = |\alpha(\omega)|^2\frac{4\Omega_0\omega}{(\Omega_0 + \omega)^2} ,
\end{equation}
which behaves very much like a probability distribution \cite{HOpaper}.  In particular it is clearly positive semi-definite, but it is also normalized in that we can show that
\begin{equation}
\label{Eq5.2}
\int_0^\infty \pi(\omega)d\omega = 1 .
\end{equation}
We then find that the properties of the dressed atom can be expressed simply in terms of $\pi(\omega)$.  In particular we find that \cite{HOpaper}
\begin{align}
\label{Eq5.3}
\langle 0|\hat{a}|0\rangle &=  0 = \langle 0|\hat{a}^\dagger|0\rangle \nonumber \\
\langle 0|(\hat{a} + \hat{a}^\dagger)^2|0\rangle &=  \Omega_0\int_0^\infty \frac{\pi(\omega)}{\omega} d\omega 
= \Omega_0\langle\langle\omega^{-1}\rangle\rangle  \nonumber \\
-\langle 0|(\hat{a} - \hat{a}^\dagger)^2|0\rangle &=   \Omega_0^{-1}\int_0^\infty \omega\pi(\omega) d\omega 
= \frac{\langle\langle\omega\rangle\rangle}{\Omega_0}  ,
\end{align}
where we have introduced the double angle bracket to denote averages over $\pi(\omega)$ so that 
\begin{align}
\label{Eq5.4}
\langle\langle\omega\rangle\rangle &=  \int_0^\infty \omega\pi(\omega) d\omega \nonumber \\
\langle\langle\omega^{-1}\rangle\rangle &=  \int_0^\infty \frac{\pi(\omega)}{\omega} d\omega .
\end{align}
The entanglement between the atom and the surrounding field means that we have the uncertainty product
\begin{equation}
\label{Eq5.5}
\Delta(\hat{a}+\hat{a}^\dagger)\Delta[-i(\hat{a}-\hat{a}^\dagger)] > 1 ,
\end{equation}
which exceeds the value (of unity) associated with the undressed, or bare, ground state of the oscillator. The increased uncertainties imply, moreover, an increased energy for the dressed ground state of the Lorentz atom:
\begin{equation}
\label{Eq5.6}
\hbar\Omega_0\langle 0|\hat{a}^\dagger\hat{a} + \frac{1}{2}|0\rangle = 
\frac{\hbar\Omega_0}{4}\left(\Omega_0\langle\langle\omega^{-1}\rangle\rangle + 
\frac{\langle\langle\omega\rangle\rangle}{\Omega_0}\right) ,
\end{equation}
which is greater that $\hbar\Omega_0/2$ for any $\pi(\omega)$\footnote{This follows by virtue of the Cauchy-Schwartz inequality, which ensures that $\langle\langle\omega^{-1}\rangle\rangle\langle\langle\omega\rangle\rangle>1$}.  This increase in the ground-state energy of the atom alone is the analogue of the behaviour found for two coupled harmonic oscillators as expressed in the inequality (\ref{Eq3.5}).  A more extended discussion of these properties may be found in \cite{HOpaper}.

\subsection{The virtual photons}

What has yet to be addressed is the properties of the electromagnetic field surrounding the Lorentz atom.  As with the atom every property of the field can be determined, at least in principle, using the exact form of the dressed operators $\hat{B}(\omega)$.  Here, however, we direct our attention to the spectrum of the virtual photons, their mean number and the correlations between the field and the atom.  To proceed we first need to express the field annihilation and creation operators in terms of the dressed operators.  To this end we write 
\begin{equation}
\label{Eq5.7}
\hat{b}(\omega) = \int_0^\infty d\omega"[\kappa(\omega,\omega")\hat{B}(\omega") + \lambda(\omega,\omega")\hat{B}^\dagger(\omega")] .
\end{equation}
Then considering the commutators $[\hat{b}(\omega'),\hat{B}(\omega)]$ and $[\hat{b}(\omega'),\hat{B}^\dagger(\omega)]$ gives
\begin{align}
\label{Eq5.8}
\kappa(\omega,\omega') &= \gamma^*(\omega',\omega) \nonumber \\
\lambda(\omega,\omega') &= -\delta(\omega',\omega)
\end{align}
so that 
\begin{equation}
\label{Eq5.9}
\hat{b}(\omega) = \int_0^\infty d\omega'[\gamma^*(\omega',\omega)\hat{B}(\omega') - \delta(\omega',\omega)\hat{B}^\dagger(\omega')] .
\end{equation}
Note the similarity with the form of $\hat{B}(\omega)$ given in (\ref{Eq4.10}) but that the arguments in the functions $\gamma$ and $\delta$ are interchanged.  The commutation relations between the bare field annihilation and creation operators implies that our functions are related by
\begin{equation}
\label{Eq5.10}
\int_0^\infty d\omega'[\gamma^*(\omega',\omega)\gamma(\omega',\nu) - \delta(\omega',\omega)\delta^*(\omega',\nu) ] = 
\delta(\omega - \nu) .
\end{equation}
Applying the same procedure to the creation and annihilation operators for the atom operators leads to the expression \cite{HOpaper}
\begin{equation}
\label{Eq5.11}
\hat{a} = \int_0^\infty d\omega[\alpha^*(\omega)\hat{B}(\omega) - \beta(\omega)\hat{B}^\dagger(\omega)] .
\end{equation}
We can use these expressions, together with the fact that the ground state is annihilated by the operators $\hat{B}(\omega)$, to examine the properties of the ground state.  In particular the spectral density of the virtual photons is
\begin{align}
\label{Eq5.12}
N(\omega) &= \langle 0|\hat{b}^\dagger(\omega)\hat{b}(\omega)|0\rangle \nonumber \\
&= \int_0^\infty d\omega' |\delta(\omega',\omega)|^2  \nonumber \\
&=\frac{V^2(\omega)\Omega_0}{4}\int_0^\infty \frac{\pi(\omega')}{(\omega'+\omega)^2\omega'}d\omega' .
\end{align}
This quantity is finite but cut-off frequency dependent for our model coupling.  The finite value follows from the general requirement that $\int_0^\infty \pi(\omega)\omega^{-1}d\omega$ must be finite in order for our diagonalization to lead to a well-behaved Hamiltonian.  We can use (\ref{Eq5.12}) to calculate the mean number of virtual photons by integrating over all $\omega$,
and also the mean energy by first multiplying this density by $\hbar\omega$ and then integrating.

The diagonalization makes it possible to calculate, also, the moments of our field creation and annihilation operators.  In particular we note that the quantities $\langle 0|\hat{b}(\omega)\hat{b}(\omega')|0\rangle$ and its complex conjugate are non-zero:
\begin{equation}
\label{Eq5.13}
\langle 0|\hat{b}(\omega)\hat{b}(\omega')|0\rangle = -\int_0^\infty d\omega"\gamma^*(\omega",\omega)\delta(\omega",\omega') .
\end{equation}
As $\hat{b}(\omega)$ commutes with $\hat{b}(\omega')$, this must be symmetric in $\omega$ and $\omega'$. These non-vanishing coherences are a reflection of the fact that the natural (virtual) excitations of the field surrounding our Lorentz atom are not simply those created by the bare field operators $\hat{b}^\dagger(\omega)$ and $\hat{b}(\omega)$. Indeed we can see this directly in the form of $\hat{b}(\omega)$ expressed in terms of $\hat{B}(\omega)$ and $\hat{B}^\dagger(\omega)$ (see Eq. (\ref{Eq5.9})).  This makes it clear that the action of $\hat{b}(\omega)$ on the true ground state, $|0\rangle$, has a finite amplitude for preparing or creating a virtual photon. This behaviour appears also in the Lorentz atom for which the mean number of (bare) excitations is \cite{HOpaper}
\begin{equation}
\label{Eq5.14}
\langle 0|\hat{a}^\dagger\hat{a}|0\rangle = \frac{1}{4}\left(\Omega_0\langle\langle\omega^{-1}\rangle\rangle + 
\frac{\langle\langle\omega\rangle\rangle}{\Omega_0} - 2\right) > 0
\end{equation}
and also in our introductory problem of two coupled oscillators introduced in section \ref{Sect3}.

	\begin{figure}[htbp] 
\centering
\includegraphics[width=8.5cm]{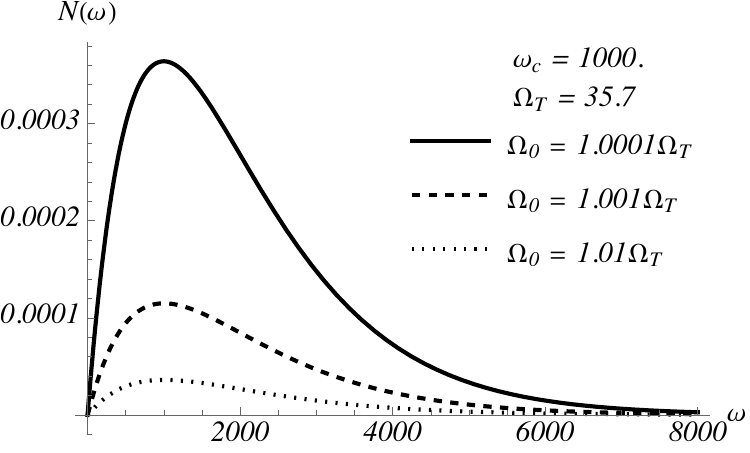}
\caption{The spectral density of the virtual photons, $N(\omega)$, as a function of the frequency $\omega$.  The frequency is expressed in units of the quantity $e^2\omega_c^2/6\pi\varepsilon_0c^3m$.} 
\label{fig:figure1}
\end{figure}

The spectral density of the virtual photons, $N(\omega)$, is plotted in Fig. \ref{fig:figure1} for three different choices of the frequency $\Omega_0$.  These are expressed in terms of the threshold frequency
\begin{equation}
\label{Eq5.14a}
\Omega_T = \int_0^\infty d\omega \frac{V^2(\omega)}{\omega} \, .
\end{equation}
This threshold frequency is the lowest frequency that allows for the diagonalisation of the Hamiltonian.  The reason for this is that the dissipation changes the oscillation frequency to $\omega_0$ given by
\begin{equation}
\label{Eq5.14b}
\omega_0^2 = \Omega_0^2 - \int_0^\infty d\omega \frac{V^2(\omega)}{\omega} 
\end{equation}
and this oscillation frequency, which appears in the Heisenberg equation of motion, must be positive \cite{HOpaper}.  

\begin{figure}[htbp] 
\centering
\includegraphics[width=8.5cm]{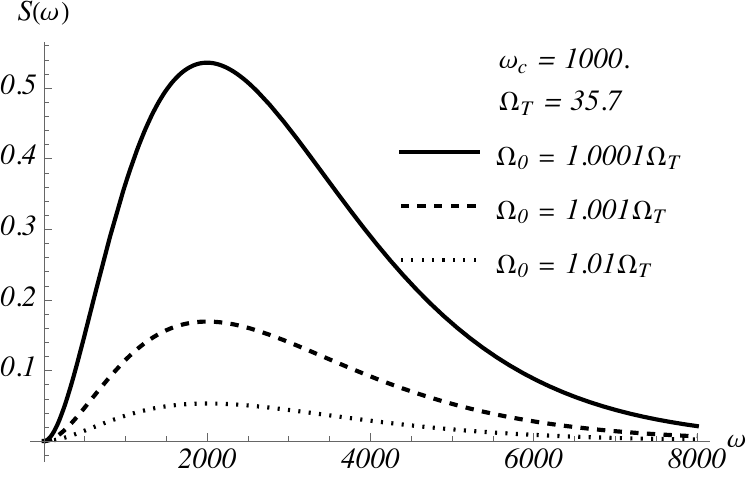}
\caption{The spectrum of the virtual photons, $S(\omega)$, as a function of the frequency $\omega$.  The frequency is expressed in units of the quantity $e^2\omega_c^2/6\pi\varepsilon_0c^3m$.} 
\label{fig:figure2}
\end{figure}

In Fig. \ref{fig:figure2} we plot the spectrum $S(\omega) = \omega N(\omega)$ for the virtual photons.  In neither of these cases is there a feature special to the frequency $\Omega_0$.  This is a simple consequence of the dominant role played by antiresonant, excitation number non-conserving, transitions in the creation of virtual photons.  We see that both $N(\omega)$ and $S(\omega)$ increase as the frequency $\Omega_0$ is reduced.  We can understand this behaviour quite simply by considering the modification of the bare ground states for the Lorentz atom and the field in lowest order perturbation theory.  The corresponding correction to this, in which the atom goes into its first excited state by emitting a photon, depends on the frequency of the emitted virtual photon, $\omega$ and on $\Omega_0$ through the factor $(\omega + \Omega_0)^{-1}$.  This modification, and the corresponding probability for the presence of a virtual photon, will \emph{increase} as the frequency $\Omega_0$ decreases.

\subsection{Atom-field correlations}
	It remains to explore the correlations between the atom and the surrounding field appearing in the ground state.  These correlations are essential in that they are associated directly with the fact that the ground state is a pure state but the states of the atom and the field alone are mixed \cite{QIbook}.  
	
	We start by observing the forms of a pair of correlations between atom field operators.  For correlations between the atomic operators and the field annihilation operators we find
\begin{widetext}
	\begin{align}
\label{Eq5.14c}
\langle 0|(\hat{a} \pm \hat{a}^\dagger)\hat{b}(\omega)|0\rangle &=  \int_0^\infty d\omega'\int_0^\infty d\omega" 
\langle 0|[\alpha^*(\omega')\hat{B}(\omega') \mp \beta^*(\omega')\hat{B}(\omega')]  
[-\delta(\omega",\omega)\hat{B}^\dagger(\omega")]|0\rangle \nonumber \\
&=  -\int_0^\infty d\omega' \delta(\omega",\omega)[\alpha^*(\omega') \mp \beta^*(\omega')] \nonumber \\
&=  -\int_0^\infty d\omega' \frac{V(\omega)}{\omega'+\omega}\frac{\Omega_0}{\omega'+ \Omega_0}|\alpha(\omega')|^2
\left[1 \mp \frac{\omega' - \Omega_0}{\omega' + \Omega_0}\right] .
\end{align}
For the correlations with the field creation operators we have
\begin{align}
\label{Eq5.15}
\langle 0|(\hat{a} \pm \hat{a}^\dagger)\hat{b}^\dagger(\omega)|0\rangle &=
\langle 0|\hat{b}^\dagger(\omega)(\hat{a} \pm \hat{a}^\dagger)|0\rangle  \nonumber \\
&= -\int_0^\infty d\omega' \frac{V(\omega)}{\omega'+\omega}\frac{\Omega_0}{\omega'+ \Omega_0}|\alpha(\omega')|^2
\left[\pm 1 - \frac{\omega' - \Omega_0}{\omega' + \Omega_0}\right] .
\end{align}
We note that all of these correlation functions are real and negative.  Combining these together we recover the 
simple expressions
	\begin{align}
\label{Eq5.16}
\langle 0|(\hat{a} + \hat{a}^\dagger)(\hat{b}(\omega) + \hat{b}^\dagger(\omega))|0\rangle 
&=  -\int_0^\infty d\omega' \frac{V(\omega)\Omega_0(\Omega_0 + \omega')}{(\omega' + \omega)^2\omega'}
\pi(\omega')  \nonumber \\
\langle 0|[-i(\hat{a} - \hat{a}^\dagger)](\hat{b}(\omega) + \hat{b}^\dagger(\omega))|0\rangle &=  0 \nonumber \\
\langle 0|(\hat{a} + \hat{a}^\dagger)[-i(\hat{b}(\omega) - \hat{b}^\dagger(\omega))]|0\rangle &=   0  \nonumber \\
\langle 0|[-i(\hat{a} + \hat{a}^\dagger)][-i(\hat{b}(\omega) - \hat{b}^\dagger(\omega))]|0\rangle 
&= \int_0^\infty d\omega' \frac{V(\omega)(\Omega_0 + \omega')}{(\omega' + \omega)^2}
\pi(\omega')  \, .
\end{align}
\end{widetext}
We note that the first of these is strictly negative and that the last one is positive.  It follows directly that the electron position and the electric field strength are anti-correlated but that the electron velocity (or momentum) is correlated with the rate of change of the electric field:
\begin{widetext}
	\begin{align}
\label{Eq5.17}
\langle 0|\hat{z}\hat{E}_z(0)|0\rangle &= \left(\frac{\hbar^2e^2}{12\pi^2\varepsilon_0c^3m\Omega_0}\right)^{1/2}
\int_0^\infty \omega^{3/2}d\omega \langle 0|(\hat{a} + \hat{a}^\dagger)(\hat{b}(\omega) + \hat{b}^\dagger(\omega))|0\rangle \nonumber \\
&= - \left(\frac{\hbar^2e^2}{12\pi^2\varepsilon_0c^3m\Omega_0}\right)^{1/2} \int_0^\infty  \omega^{3/2}V(\omega)d\omega
\int_0^\infty d\omega' \frac{\Omega_0(\Omega_0 + \omega')}{(\omega' + \omega)^2\omega'}
\pi(\omega')  \nonumber \\
\langle 0|\frac{d\hat{z}}{dt}\frac{d\hat{E}_z(0)}{dt}|0\rangle 
&= \left(\frac{\hbar^2e^2\Omega_0}{12\pi^2\varepsilon_0c^3m}\right)^{1/2} 
\int_0^\infty \omega^{5/2}V(\omega)d\omega \int_0^\infty d\omega'
\frac{(\Omega_0 + \omega')}{(\omega' + \omega)^2}
\pi(\omega') .
\end{align}
\end{widetext}
It is possible to regularize these correlation functions using the procedure utilized in the preceding section, but the present form suffices to understand their nature. The fact that the signs of these correlation functions differ from the position and momentum correlations found for the coupled oscillators in section \ref{Sect3} stems from the fact that the interaction terms have opposite signs and this, in turn, is a consequence of the negative charge of the electron.  It is straightforward to make physical sense of these correlations: if the fluctuating vacuum electric field points in the positive $z$ direction then the electron is pushed towards negative values of $z$ and similarly a fluctuation on the negative $z$ direction will tend to move the electron towards positive values of $z$.  Similarly, a motion of the electron towards greater values of $z$ will be associated with a decrease of the electric field in the positive $z$ direction (or an increase of the electric field in the negative $z$ direction).

\subsection{Characteristic functional and general properties}

The ground state is fully characterised by the normal-ordered moments of the creation and annihilation operators for the Lorentz atom and the field and it is convenient to have a simple way to calculate these.  To this end we can compute the normal-ordered characteristic functional. This is the extension of the familiar characteristic function for a discrete oscillator \cite{PaulR} generalised to include the properties of the continuum field.  To this end we define the characteristic functional to be
\begin{multline}
\label{Eq5.18}
\chi[\xi(\omega),\eta] =\\
 \langle 0|e^{\int d\omega \xi(\omega) \hat{b}^\dagger(\omega)}
e^{\eta\hat{a}^\dagger}e^{-\eta^*\hat{a}}e^{-\int d\omega'\xi^*(\omega')\hat{b}(\omega')}|0\rangle \, .
\end{multline}
From this we can determine any desired, normal-ordered expectation value by a combination of partial differentiation and functional differentiation \cite{PaulR,Rivers,Zinn-Justin,Mandl}.  
For example
\begin{multline}
\label{Eq5.19}
\langle 0|\hat{b}^\dagger(\nu)\hat{a}^\dagger\hat{a}\hat{b}(\nu')|0\rangle 
= \\
\left. \frac{\delta}{\delta \xi(\nu)}\cdot\frac{\partial}{\partial \eta} \cdot -\frac{\partial}{\partial \eta^*}\cdot -\frac{\delta}{\delta \xi^*(\nu')}
\chi[\xi(\omega),\eta]\right|_{\xi)\omega),\eta = 0} \, .
\end{multline}
It is straightforward to calculate our characteristic functional by first writing the annihilation and creation operators for the Lorentz atom and for the field in terms of the dressed operators, $\hat{B}(\omega)$ and $\hat{B}^\dagger(\omega')$, and then applying a disentangling theorem to normal order these \cite{PaulR}.  The result is
\begin{multline}
\label{Eq5.20}
\chi[\xi(\omega),\eta] = \\
e^{-\frac{1}{2}\int d\nu q(\nu)p(\nu)}e^{-\frac{1}{2}\int d\nu' q^*(\nu')p^*(\nu')}
e^{-\int d\nu" p^*(\nu")p(nu")} \, ,
\end{multline}
where
\begin{align}
\label{Eq5.21}
p(\nu) &= \eta\beta^*(\nu) + \int d\omega \xi(\nu) \delta^*(\nu,\omega)  \nonumber \\
q(\nu) &= \eta\alpha(\nu) + \int d\omega \xi(\omega) \gamma(\nu,\omega) \, .
\end{align}
We note that the characteristic functional is Gaussian in form, peaked around $\eta = 0 = \xi(\omega)$.  Such pure states are a generalisation of the squeezed states familiar from quantum optics \cite{PaulR,Loudon}.  Such states, characterised by a Gaussian Wigner function, are referred to as Gaussons \cite{BBAnnPhys,BBGlauberBook,GordonY}.

A few simple examples may serve to illustrate the general method:
\begin{align}
\label{Eq5.22}
\langle 0|\hat{a}^2|0\rangle &=  \left.\frac{\partial^2}{\partial \eta^{*2}}\chi[\xi(\omega),\eta]\right|_{\xi(\omega),\eta=0}
\nonumber \\
&=  \int d\omega |\beta(\omega)|^2  \nonumber \\
&=  -\frac{1}{4}\left(\frac{\langle\langle\omega\rangle\rangle}{\Omega_0} - 
\Omega\langle\langle\omega^{-1}\rangle\rangle \right)  \nonumber \\
\langle 0|\hat{b}^\dagger(\nu)\hat{b}(\nu')|0\rangle &=  \left. -\frac{\delta}{\delta\xi(\nu)}
\frac{\delta}{\delta\xi^*(\nu')}\chi[\xi(\omega),\eta]\right|_{\xi(\omega),\eta=0} \nonumber \\
&=  \int d\omega \delta^*(\omega,\nu)\delta(\omega,\nu') \nonumber \\
\langle 0|\hat{a}\hat{b}(\nu)|0\rangle &=  \left. \frac{\partial}{\partial \eta^*}
\frac{\delta}{\delta\xi^*(\nu)}\chi[\xi(\omega),\eta]\right|_{\xi(\omega),\eta=0}  \nonumber \\
&=  -\frac{1}{2}\int d\omega\left(\gamma^*(\omega,\nu)\beta(\omega) + \delta(\omega,\nu)\alpha^*(\omega)\right) \, .
\end{align}
This method also reproduces the correlation functions and moments given in the preceding subsections.

\section{Conclusion}

The Lorentz model of an atom treats an atomic electron as if it were harmonically bound rather than by the Coulomb potential.  Its simplicity means that it is commonly used in treatments of linear optical processes in material media \cite{Jackson,Bleaney,Dressel}.  We have extended this idea to treat the dressing of an atomic electron and, in particular, to model the properties of the dressed ground state.  The advantage of this model is that it allows a fully analytical derivation of the properties of both the Lorentz atom and also of the surrounding virtual photons.  That this is possible is a direct consequence of the fact that we are then treating the coupling between a harmonically bound electron and the surrounding bosonic field.  Within the electric dipole approximation, this reduces, effectively, to the coupling between a discrete oscillator (the atom) and a single continuum characterized by its frequency.  This model is amenable to an exact diagonalization and hence to an exact ground state that is an entangled state of the atom and field \cite{HOpaper}.

Our analysis led to expressions for the mean excitations of the atom and of the electromagnetic field associated with the dressed ground state.  It also revealed correlations between the position and motion of the electron with the local fluctuating vacuum field.  The form of these, or at least their signs fits naturally with the expected response of a bound electron to a fluctuating electric field.

It should be stated clearly that real atoms are rather different, and certainly more complicated in their structure than a Lorentz atom.  This means that we do not expect real atoms to be behave quantitively in the manner explored here.  Nevertheless, we can expect many of the qualitative features to occur, such at the correlations between the displacement of the electrons from the nucleus and the vacuum electric field.  Indeed such simplified reasoning has been helpful in the past, not least in Welton's derivation of the Lamb shift \cite{Welton}.  We hope that the model introduced here may similarly aid physical reasoning in the further study of virtual photons and of their significance.

\acknowledgments{It is a pleasure to acknowledge the debt that we owe to Prof. Persico and also to Profs. Power and Thirunamachandran, who did so much to develop the field of non-relativistic quantum electrodynamics, including the crucial role played by dressed states and the virtual photons they include. This work was supported by the Royal Society, grant number RP150122. All of the authors contributed to the research described in this paper.  SMB drafted the original manuscript, which was amended and edited following comments from JDC and SC. The authors declare no conflict of interest.
}

\appendix
\section{Derivation of the interaction term in the Hamiltonian (\ref{Eq4.1})}
\label{AppA}

In order to derive our model Hamiltonian we need to reduce the full electric field operator into an effective one dimensional form in which the field annihilation and creation operators are labeled only by the corresponding temporal frequency.  That this is possible is a consequence of the electric-dipole approximation as we need the electric field operator only at a single point, which we take to be the origin.  

We start with the expression, in the Schr\"{o}dinger picture, for the electric field operator written in terms of a set of discrete modes \cite{QTL}:
\begin{equation}
\label{EqA.1}
\hat{\bf E}({\bf r}) = \sum_{\bf k}\sum_\lambda {\bf e}_{{\bf k},\lambda}  \left(\frac{\hbar\omega_{\bf k}}{2\varepsilon_0\mathcal{V}}\right)^{1/2}
\left[i\hat{a}_{{\bf k},\lambda}e^{i{\bf k}\cdot{\bf r}} - i\hat{a}^\dagger_{{\bf k},\lambda}e^{-i{\bf k}\cdot{\bf r}} \right] .
\end{equation}
We choose the wavevectors and polarizations to have cartesian forms as follows
\begin{align}
\label{EqA.2}
{\bf k} &=  k(\sin\theta_{\bf k}\cos\phi_{\bf k},\sin\theta_{\bf k}\sin\phi_{\bf k},\cos\theta_{\bf k}) \nonumber \\
{\bf e}_{{\bf k},1} &=  (\sin\phi_{\bf k},-\cos\phi_{\bf k},0)  \nonumber \\
{\bf e}_{{\bf k},2} &=  (\cos\theta_{\bf k}\cos\phi_{\bf k},\cos\theta_{\bf k}\sin\phi_{\bf k},-\sin\theta_{\bf k}) ,
\end{align}
so that the polarization vectors are normalized and mutually orthogonal and they are also orthogonal to the wavevector ${\bf k}$.  We note that with this choice the ${\bf e}_{{\bf k},1} $ are all orthogonal to $\hat{z}$ and so we need only consider the second polarization for which
\begin{equation}
\label{EqA.3}
\hat{\bf z}\cdot{\bf e}_{{\bf k},2} = -\sin\theta_{\bf k} .
\end{equation}
It follows that we can drop the second polarization and write the required $z$-component of the electric field 
in the form
\begin{equation}
\label{EqA.4}
\hat E_z({\bf r}) = -i\sum_{\bf k}  \left(\frac{\hbar\omega_{\bf k}}{2\varepsilon_0\mathcal{V}}\right)^{1/2}
\sin\theta_{\bf k} \left[\hat{a}_{\bf k}e^{i{\bf k}\cdot{\bf r}} - \hat{a}^\dagger_{\bf k}e^{-i{\bf k}\cdot{\bf r}} \right] .
\end{equation}
Next we take the continuum (infinite volume) limit \cite{QTL},
\begin{equation}
\label{EqA.5}
\sum_{\bf k} \rightarrow \frac{\mathcal{V}}{(2\pi)^3}\int d^3k ,
\end{equation}
together with the continuum annihilation and creation operators:
\begin{equation}
\label{EqA.6}
\hat{a}_{\bf k} \rightarrow \hat{a}({\bf k}) \sqrt{\frac{(2\pi)^3}{\mathcal{V}}}.
\end{equation}
With these changes, our $z$-component of the electric field operator becomes
\begin{widetext}
	\begin{equation}
\label{EqA.7}
\hat E_z({\bf r}) = -\frac{1}{(2\pi)^{3/2}}\int d^3k \sin\theta_{\bf k}
\left(\frac{\hbar\omega_{\bf k}}{2\varepsilon_0}\right)^{1/2}
i\left[\hat{a}({\bf k})e^{i{\bf k}\cdot{\bf r}} - \hat{a}^\dagger(\bf k)e^{-i{\bf k}\cdot{\bf r}} \right] .
\end{equation}
\end{widetext}
We seek to express this as a single integral over $\hat{a}(\omega)$.  To this end we define
\begin{equation}
\label{EqA.8}
\hat{a}(k) = k\sqrt{\frac{3}{8\pi}}\int_0^{2\pi}d\phi_{\bf k}\int_0^\pi \sin^2\theta_{\bf k} d\theta_{\bf k}\hat{a}({\bf k}),
\end{equation}
where the normalization has been chosen so that $[\hat{a}(k),\hat{a}^\dagger(k')] = \delta(k - k')$.  Finally, we would like an integral in terms of the frequency, $\omega$.  To this end we note that $k = \omega/c$ and tidy up the factor $i$ by writing 
\begin{equation}
\label{EqA.9}
\hat{a}(k) = i\sqrt{c}\hat{b}(\omega) ,
\end{equation}
so that the $z$-component of the electric field operator at the origin is
\begin{equation}
\label{EqA.10}
\hat{E}_z(0) = \frac{1}{\sqrt{3}\pi} \int_0^\infty \omega^{3/2}\left(\frac{\hbar}{2\varepsilon_0c^3}\right)^{1/2}
[\hat{b}(\omega) + \hat{b}^\dagger(\omega)] d\omega .
\end{equation}
The product of this with the operator $\hat{z}$ corresponding to the position of the electron and the electron charge, $-e$, gives the electric-dipole interaction term in our Hamiltonian. We emphasize that this treatment is {\em exact} within the electric-dipole approximation.

\section{Diagonalizing the Hamiltonian}
\label{AppB}

We seek to diagonalize the Hamiltonian (\ref{Eq4.5}), by which we mean rewriting it in the form of a continuum of uncoupled, or dressed, operators:
\begin{equation}
\label{EqB.1}
\hat{H} = \int_0^\infty \hbar{\omega} \: \hat{B}^\dagger(\omega)\hat{B}(\omega) ,
\end{equation}
where we have omitted a physically insignificant constant term.  Our approach is generalization \cite{HOpaper,Bruno} of one developed by Fano to treat configuration interactions in atomic physics \cite{PaulR,Fano}.

We proceed by writing the dressed annihilation operators as a linear combination of the (undressed) operators for the Lorentz atom and the field:
\begin{equation}
\label{EqB.2}
\hat{B}(\omega) = \alpha(\omega)\hat{a} + \beta(\omega)\hat{a}^\dagger 
+\int_0^\infty d\omega'[\gamma(\omega,\omega')\hat{b}(\omega') + \delta(\omega,\omega')\hat{b}^\dagger(\omega)] .
\end{equation}
The dressed annihilation operators are required to satisfy the operator eigenvalue equation (\ref{Eq4.9}).  Enforcing this and comparing the coefficients of the diverse operators leads to the coupled equations
\begin{align}
\label{EqB.3}
\alpha(\omega)\Omega_0 + \frac{1}{2}\int_0^\infty d\omega' V(\omega')[\gamma(\omega,\omega') - \delta(\omega,\omega')]
&= \alpha(\omega)\omega \nonumber \\
-\beta(\omega)\Omega_0 +  \frac{1}{2}\int_0^\infty d\omega' V(\omega')[\gamma(\omega,\omega') - \delta(\omega,\omega')]
&= \beta(\omega)\omega \nonumber \\
\frac{1}{2}V(\omega')[\alpha(\omega) - \beta(\omega)] + \gamma(\omega,\omega')\omega' &= \gamma(\omega,\omega')\omega
\nonumber \\
\frac{1}{2}V(\omega')[\alpha(\omega) - \beta(\omega)] - \delta(\omega,\omega')\omega' &= \delta(\omega,\omega')\omega .
\end{align}
Solving these, with the aid of a method introduced by Dirac \cite{Dirac} leads to the expressions (\ref{Eq4.11}). Finally, imposing the bosonic commutation relation
\begin{equation}
\label{EqB.4}
[\hat{B}(\omega),\hat{B}^\dagger(\omega')] = \delta(\omega-\omega')
\end{equation}
leads to an expression for $|\alpha(\omega)|^2$:
\begin{equation}
\label{EqB.5}
|\alpha(\omega)|^2 = \frac{(\omega + \Omega_0)^2}{\Omega^2_0V^2(\omega)}\left(\frac{1}{Y^2(\omega) + \pi^2}\right) .
\end{equation}
Further details may be found in \cite{HOpaper,Bruno}.

\end{document}